\documentclass[reprint,a4paper,australian,amsmath,aps,prd,twocolumn,superscriptaddress,preprintnumbers,noshowpacs]{revtex4-1}
\usepackage{CJK}
\usepackage{amsmath}
\usepackage{enumerate}
\usepackage{amssymb}
\usepackage{amsfonts}
\usepackage{mathrsfs}
\usepackage{latexsym}
\usepackage{bm}
\usepackage{amscd}
\usepackage{graphicx}
\usepackage{epsfig}
\usepackage{hhline,multirow}
\usepackage{dcolumn}
\usepackage{url}
\usepackage[dvipsnames]{xcolor}
\definecolor{DarkBlue}{rgb}{0.0, 0.0, 0.4}
\usepackage[colorlinks=true,linkcolor=blue,urlcolor=DarkBlue]{hyperref}
\usepackage{indentfirst}
\usepackage{subfigure}
\usepackage{psfrag}
\usepackage{slashed}
\usepackage{float}
\usepackage{setspace}

\newcommand{\mpi}{M_\pi}

\begin{document}

\preprint{ADP-20-28/T1138}

\title{Chiral extrapolation of the magnetic polarizability of the neutral pion}
\author{Fangcheng He}
\affiliation{Institute of High Energy Physics, CAS,
	Beijing 100049, China}
\affiliation{CAS Key Laboratory of Theoretical Physics, Institute of Theoretical Physics, CAS, Beijing 100190, China}
\author{D. B. Leinweber}
\affiliation{Centre for the Subatomic Structure of Matter (CSSM), Department of Physics, University of Adelaide,
	Adelaide SA 5005, Australia}
\author{A. W. Thomas}
\affiliation{CoEPP and CSSM, Department of Physics, University of Adelaide,
	Adelaide SA 5005, Australia}
\author{P. Wang}
\affiliation{Institute of High Energy Physics, CAS,
	Beijing 100049, China}
\affiliation{Theoretical Physics Center for Science Facilities, CAS,
	Beijing 100049, China}

\begin{abstract}

The magnetic polarizability of the neutral pion has been calculated in the background
magnetic-field formalism of Lattice QCD.  In this investigation, the chiral extrapolation of these
lattice results is considered in a formalism preserving the exact leading nonanalytic terms of
chiral perturbation theory.  The $n_f = 2 + 1$ numerical simulations are electro-quenched, such
that the virtual sea-quarks of the QCD vacuum do not interact with the background field.  To
understand the impact of this, we draw on partially quenched chiral perturbation theory and
identify the leading contributions of quark-flow connected and disconnected diagrams.  While
electro-quenching does not impact the leading-loop contribution to the magnetic polarizability, the
loops which generate the leading term have yet to be considered in lattice QCD simulations.
Lattice QCD results are used to constrain the analytic terms in the chiral expansion and
supplementing those with the two-loop result from chiral perturbation theory enables an evaluation
of the polarizability at the physical quark mass.  The resulting magnetic polarizability of the
neutral pion is $\beta_{\pi^0}=3.44(19)^{\rm stat}(37)^{\rm syst}\times 10^{-4}$ fm$^3$, which lies
just above the $1 \sigma$ error bound of the experimental measurement.
\end{abstract}


\maketitle

\section{Introduction}

The electromagnetic polarizabilities of hadrons provide important insights into the structure of
hadrons related to their response to electromagnetic fields.  The polarizabilities are manifest in the
shape of the $\gamma$-hadron Compton scattering angular distribution.  They provide an interesting
forum for the confrontation of experiment and theoretical approaches, challenging the current
understanding of hadron structure and generating new insights into the essential mechanisms of
Quantum Chromodynamics (QCD) in the low-energy regime.

Herein, our focus is on the lightest hadron, the pion.  Experimentally, pion electromagnetic
polarizabilities have been extracted from radiative pion photoproduction \cite{Aibergenov:1986gi},
pion nucleus scattering \cite{Antipov:1982kz,Antipov:1984ez} and from the cross section of the
$\gamma\gamma\rightarrow \pi\pi$ process \cite{Boyer:1990vu,Marsiske:1990hx,Filkov:2005ccw}.  On
the other hand, many theoretical approaches have been considered in understanding pion
polarizabilities, including quark models \cite{Bernard:1988wi,Ivanov:1991kw,Bernard:1992mp}, the
bosonized NJL model \cite{Wilmot:2002gg}, chiral perturbation theory
\cite{Bijnens:1987dc,Bellucci:1994eb,Burgi:1996qi,Gasser:2005ud,Moinester:2019sew}, dispersion sum
rules \cite{Filkov:1982cx,Donoghue:1993kw} and the linear sigma model \cite{Bernard:1988gp}.

The most rigorous formalism for the study of QCD in the low-energy regime is lattice gauge theory.
Here, spacetime is discretised onto a finite-volume lattice enabling numerical simulations on
supercomputers.  While the introduction of nonperturbatively-improved lattice gauge and fermion
actions have enabled excellent control of the discretisation errors, finite-volume effects and
quark-mass extrapolations/interpolations are quantified through the formalism of chiral effective
field theory.  This is the focus of the current investigation.

To compare lattice QCD results with experiment, one considers corrections associated with the
finite-volume of the lattice, extrapolates/interpolates lattice results typically at several input
quark masses to the physical point and finally accounts for any missing contributions.  The latter
are often associated with the neglect of quark-flow disconnected diagrams in the lattice QCD
simulations, due to the numerical difficulty in obtaining precise estimates.  For the magnetic
polarizability under consideration herein, the sea-quarks in disconnected loops are effectively
charge neutral and the calculations are said to be electro-quenched.

Recently, the formalism of lattice QCD in the presence of a uniform background magnetic field
\cite{Burkardt:1996vb} has been used to calculate the magnetic polarizability of the nucleon and
pion
 \cite{%
Primer:2013pva,
Bignell:2018acn,
Bignell:2020xkf,
Luschevskaya:2014lga,
Luschevskaya:2015cko,
Bignell:2019vpy,
Ding:2020hxw,
Bignell:2020dze
}.  
While the chiral extrapolation of the nucleon magnetic polarizability has been considered
\cite{Hall:2013dva, Bignell:2018acn,Bignell:2020xkf}, a chiral extrapolation of lattice QCD results
for the neutral pion magnetic polarizability remains.  In this article, we will extrapolate the
lattice QCD results of Ref.~\cite{Bignell:2020dze} for the magnetic polarizability of the neutral
pion, $\beta_{\pi^0}$, to the physical pion mass.  These results employ a new Laplacian-mode
projection technique that isolates the state of interest and enables accurate determinations of the
small energy shifts induced by the background magnetic field.  We will draw on partially quenched
chiral perturbation theory to identify the leading contributions of quark-flow connected and
disconnected diagrams separately and include the contributions of the missing terms.

The pion-photon scattering amplitude is first considered at the one-loop level in
partially-quenched chiral perturbation theory.  Remarkably the structure of the four-pion vertex
causes the sea-quark loop contributions to the magnetic polarizability at one loop to vanish. Thus
the fact that the lattice simulations are electro-quenched has no impact on the one
loop-contributions to the magnetic polarizabilities.  The origin of the one-loop contributions is
associated with the quark-annihilation contractions of the quark field operators of the
neutral-pion interpolating fields.

The one-loop diagram provides a leading model-independent constant term in the expansion of the
Compton amplitude.  Because the magnetic polarizability contribution to the Compton scattering
amplitude is typically written in terms of $\mpi$ times the magnetic polarizability,
$\beta_{\pi^0}$, the expansion of the latter in terms of $\mpi$ starts at order $1/\mpi$ governed
by the aforementioned model-independent constant, followed by odd powers of $\mpi$. We do not refer
to these terms as non-analytic because in the expansion of the Compton amplitude they correspond to
integer powers of $\mpi^2 \propto m_q$. The leading non-analytic behaviour first occurs in two-loop
chiral perturbation theory though the appearance of logarithms of $\mpi$.

We find that the lattice QCD results for $\beta_{\pi^0}$ are described very well over the available
pion-mass range by an expansion in powers of $\mpi$ involving three terms.  Upon adding the loop
contributions missing in the current lattice simulations at the physical pion mass, we find that
the magnetic polarizability of the neutral pion is $\beta_{\pi^0} = 3.44(19)(37)\times 10^{-4}$
fm$^3$, lying just above the $1 \sigma$ error bound of the experimental measurement.

The paper is organized in the following way. In Sec.~II, we review the magnetic polarizability of
the neutral pion in the context of lattice QCD, using partially-quenched chiral effective theory.
Numerical results of the chiral extrapolation are presented in Sec.~III and Sec.~IV provides a
summary.

\section{Magnetic polarizability of the neutral pion}
\begin{figure}[t!]
\begin{center}
\includegraphics[width=0.6\columnwidth]{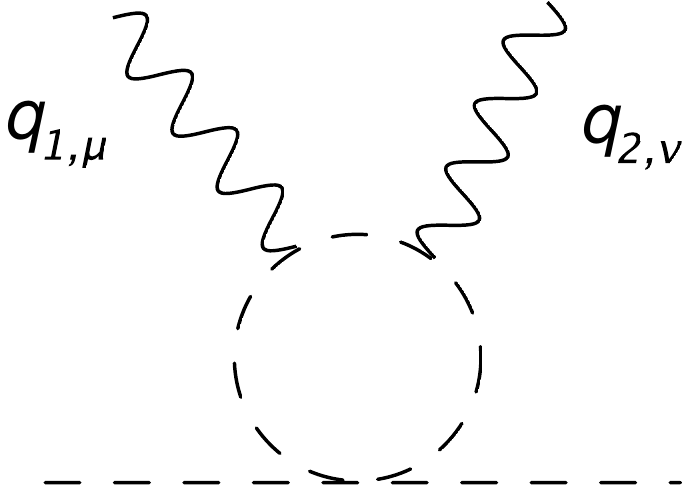}
\caption{The leading one-loop diagram for the pion magnetic polarizability.  Both $\pi^0 \pi^0
  \pi^+ \pi^+$ and $\pi^0 \pi^0 \pi^- \pi^-$ vertices contribute with the same sign.  }
\label{fig1}
\end{center}
\end{figure}
For pion-photon scattering, the Taylor expansion of the Compton amplitude 
in photon energies at threshold can be expressed as
\begin{eqnarray}
\label{mag}
T&=&-2\, \left [ \vec{\epsilon}_1\cdot\vec{\epsilon}_2^{\,*}\, (e^2 - 4\pi\,
  M_\pi\alpha_\pi \, \omega_1 \, \omega_2)- \right . \nonumber \\
&&\qquad \left . 4\pi \,
  M_\pi \, \beta_\pi\, ( \vec{q}_1\times\vec{\epsilon}_1) \cdot
  (\vec{q}_2\times\vec{\epsilon}_2^{\,*})+ \cdots \right ] \, ,
\end{eqnarray}
where $\alpha_\pi$ and $\beta_\pi$ are the electric and magnetic polarizabilities respectively.
There have been several calculations of the pion electromagnetic polarizabilities in chiral perturbation
theory \cite{Bijnens:1987dc,Bellucci:1994eb,Burgi:1996qi,Gasser:2005ud,Moinester:2019sew}.  The chiral Lagrangian
is composed of the following terms having different chiral orders.
\begin{eqnarray}
\label{eq:Lagrangian}
\mathcal{L}=\mathcal{L}_2+\mathcal{L}_4+\mathcal{L}_6+\cdots \, ,
\end{eqnarray}
where the subscripts refer to the chiral order.  The expression for $\mathcal{L}_2$ is 
\begin{eqnarray}
\mathcal{L}_2=\frac{F_\pi^2}{4}\, Tr\left [ D_\mu\, U\, D^\mu\, U^\dag \right ] +
\frac{F_\pi^2}{4}\, Tr\left [ m \left ( U+U^\dag \right ) \right ] \, ,
\label{eq:Lag}
\end{eqnarray}
where $U=e^{2i\phi/F_\pi}$, $\phi$ is the matrix of pseudoscalar fields
\begin{eqnarray}
\phi=\frac1{\sqrt{2}}\left(                
\begin{array}{lcr}                       
\frac1{\sqrt{2}}\pi^0+\frac1{\sqrt{6}}\eta & \pi^+ & K^+ \\    
~~\pi^- & -\frac1{\sqrt{2}}\pi^0+\frac1{\sqrt{6}}\eta & K^0 \\    
~~K^- & \bar{K}^0 & -\frac2{\sqrt{6}}\eta
\end{array}             
\right),
\end{eqnarray}
and $m$ is the quark mass matrix expressed as
\begin{eqnarray}
m=\left(                
\begin{array}{ccc}                       
M_\pi^2 & 0  &0  \\    
~0 & M_\pi^2 &0 \\    
~0 & 0 & 2M_K^2-M_\pi^2
\end{array}             
\right).
\end{eqnarray}
The one-loop Feynman diagram for the pion magnetic polarizability is illustrated in Fig.~1.

For the neutral pion, the scattering amplitude is written as 
\begin{widetext}
\begin{eqnarray}\label{full}
T=\frac{-ie^2}{3F_\pi^2}\, \int \frac{d^4k}{(2\pi)^4} \,
\frac{[\,2\,(k-q_1)\cdot (k-q_2)+M_\pi^2]\,(2k-q_2)_\nu\,(2k-q_1)_\mu}{(k^2-M_\pi^2)\,
(\,(k-q_2)^2-M_\pi^2)\,(\,(k-q_1)^2-M_\pi^2)}\,
\epsilon^\mu(q_1)\,\epsilon^{\nu*}(q_2) + C.S.
\end{eqnarray}
\end{widetext}
where $C.S.$ denotes crossing symmetry where the photons labelled $q_{1,\mu}$ and $q_{2,\nu}$ in
Fig.~1 couple with the opposite time ordering. Here, the $\pi^0 \pi^0 \pi^+ \pi^+$ vertex is
considered.  The same result is obtained for the $\pi^0 \pi^0 \pi^- \pi^-$ vertex and the full
result contains both contributions.

This one-loop diagram generates a leading model-independent constant term in the expansion of the
Compton amplitude, thus providing a leading divergent term in the chiral expansion of the magnetic
polarizability
\begin{equation}
\beta_{\pi^0} = \frac{\alpha}{32\, \pi^2\, F_\pi^2\, \mpi }\, \left ( \frac{1}{3} + {\cal O}(\mpi^2)
  \right ) \, .
\label{eq:leading}
\end{equation}

In order to appreciate what is included in current lattice QCD simulations in which photon coupling
to disconnected quark loops is not included, we now consider the separation of the valence and loop
contributions to $\beta_{\pi^0}$ in partially-quenched chiral perturbation theory.  This was first
considered by Hu {\it et al.} \cite{Hu:2007ts} in the graded symmetry formalism
\cite{Bernard:1992mk} at one loop.  Here we briefly review these results in the complementary
diagrammatic formalism \cite{Leinweber:2002qb,Hall:2013dva}.

Considering the diagrammatic approach, all the quark-flow diagrams for the $\pi^+,\ \pi^0$ and
$\pi^-$ dressings of the neutral pion are illustrated in Fig.~2.  As we are applying the formalism
to $n_f = 2 + 1$ dynamical-fermion simulations, we do not consider the additional quark flows
associated with the flavour-singlet $\eta'$ meson \cite{Hu:2007ts}, as there are no partial
quenching effects to consider and it remains massive $\sim 1$ GeV.

Figures 2(a) through 2(d) include sea-quark-loop contributions and because they only involve the
$u$ and $d$ flavors, the contribution of these sea-quark-loop diagrams can be isolated in the
diagrammatic approach by replacing the light sea-quark-loop flavor with a strange sea-quark-loop
flavor \cite{Leinweber:2002qb,Hall:2013dva}. Thus, they can be calculated through the consideration
of a $K$-meson loop with the $K$-meson mass replaced by pion mass.

The Compton scattering amplitude for the average of Figs.~2(b) and (c) composing the $\pi^+$
dressing of the neutral pion can be expressed as
\begin{widetext}
\begin{eqnarray}
T=\frac{-ie^2}{6F_\pi^2}\,\int \frac{d^4k}{(2\pi)^4}\, 
\frac{[(k-q_1)\cdot(k-q_2)-M_\pi^2]\, (2k-q_2)_\nu\, (2k-q_1)_\mu}{(k^2-M_\pi^2)\,
  (\,(k-q_2)^2-M_\pi^2)\,(\,(k-q_1)^2-M_\pi^2)}\, 
\epsilon^\mu(q_1)\, \epsilon^{\nu*}(q_2) + C.S. 
\end{eqnarray}
\end{widetext}
From the above equation, one can see that not only does the coefficient differ from that in
Eq.~(\ref{full}) but the structure is also different.  There is a sign change for the $M_\pi^2$
term in the numerator.  This leads to an exact cancelation of the two contributions to $\beta_\pi$
after the integral over $k$ has been carried out.

The origin of this cancelation is in a reduction of the contribution of the four-meson vertex with
two-derivatives from the Lagrangian of Eq.~(\ref{eq:Lag}) by a factor of four for the $K$-meson
loop.  In contrast, the four-meson vertex of the mass insertion remains the same, thus generating a
new cancelation.  As a result, this quark flow does not generate the structure of the
$\beta_{\pi}$ term in Eq.~(\ref{mag}).  The situation is the same for the $\pi^-$ dressings of
Figs.~2(a) and (d) and the $\pi^0$ dressings of Figs.~2(a) through (d).  As a result, Figs.~2(a)
through (d) do not contribute to $\beta_{\pi^0}$.

Thus the fact that the lattice simulations are electro-quenched has no impact on the leading
one-loop contribution to the magnetic polarizability. Although the charge on the quark loop is
set to zero in the lattice QCD simulations, the vanishing of this quark flow prevents the
electro-quenched approximation from impacting the leading one-loop contribution.

\begin{figure}[t]
\begin{center}
\includegraphics[width=0.8\columnwidth]{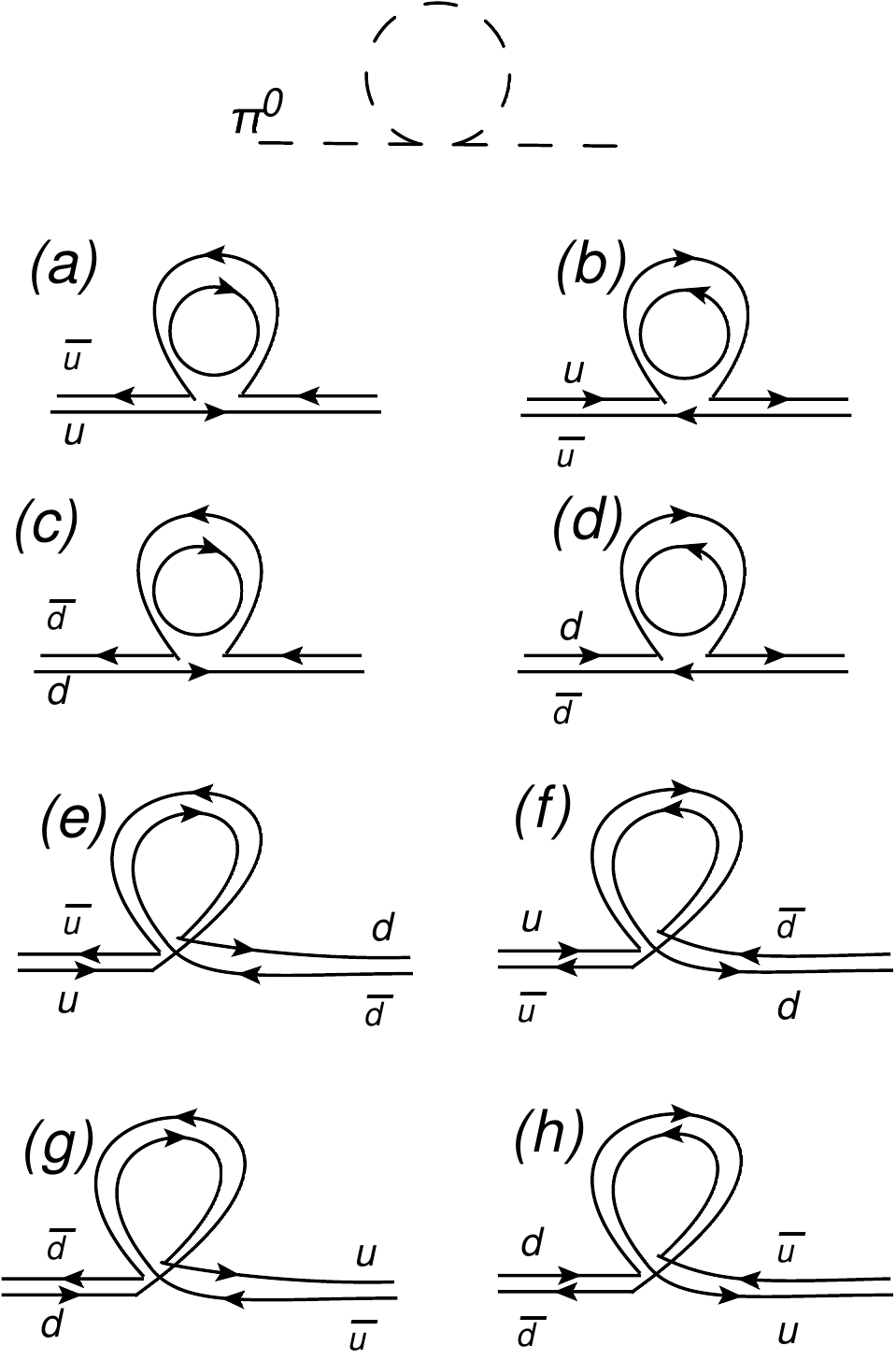}
\caption{ 
  One loop quark-flow diagrams for the $\pi^+,\ \pi^0$ and $\pi^-$ dressings of the neutral pion.
  The leading contribution to the magnetic polarizability is obtained by attaching two photons to
  the quark-flow lines of the meson loop in four different ways, outside-outside, inside-inside and
  the two inside-outside possibilities.  Diagrams (a) through (d) include sea-quark-loop
  contributions.  As the lattice results are electro-quenched, photon couplings to the inside lines
  are not included.  However, this vertex involving a sea-quark-loop does not contribute to the
  magnetic polarizability, as described in the text.  Diagrams (e) through (h) are
  quark-annihilation contractions of the quark field operators of the neutral-pion interpolating
  fields.  These quark-flow connected loop diagrams remain to be calculated in lattice QCD and are
  not included in the simulation results of Ref.~\cite{Bignell:2020dze}.
}
\label{fig2}
\end{center}
\end{figure}

It follows from the preceding discussion that only the diagrams of Figs.~2(e) through Fig.~2(h)
contribute to the $\pi^0$ magnetic polarizability \cite{Hu:2007ts}.

However, the quark-annihilation loop diagrams of Figs.~2(e) through (h) have yet to be calculated in
lattice QCD and are not included in the lattice simulation results of Ref.~\cite{Bignell:2020dze}.
Thus, the lattice QCD results which we analyze here correspond to the tree level contribution in
effective field theory.

Tree-level contributions to the Compton amplitude are analytic in the quark mass $\propto \mpi^2$.
Because the magnetic polarizability contribution to the Compton scattering amplitude is
proportional to $\mpi\, \beta_{\pi^0}$, the tree-level expansion of $\beta_{\pi^0}$ starts at order $1/\mpi$
with the form
\begin{eqnarray}\label{fit0}
\beta_{\pi^0}^\text{tree}&=&a_{-1}\, \frac{1}{\mpi} + a_1\, M_\pi + a_3 \, M_\pi^3 + \cdots \,.
\label{eq:tree}
\end{eqnarray}
As highlighted in Eq.~(\ref{eq:leading}) the one-loop diagram of Fig.~1 generates a
model-independent contribution at the leading order of $1/\mpi$.  However, tree-level physics
can also contribute.

\begin{figure}[b]
\begin{center}
\includegraphics[width=\columnwidth]{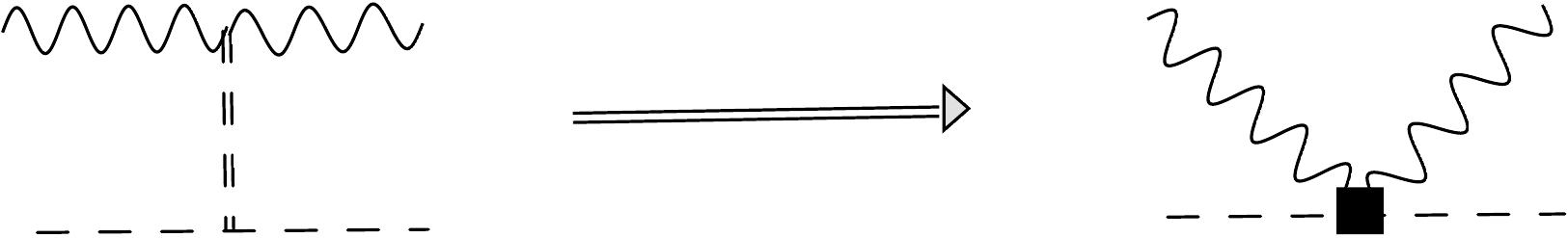}
\caption{The $\sigma$ exchange channel. Double, dashed and wavy lines represent the $\sigma$-meson,
  pion and photon, respectively.}
\label{fig:sigma}
\end{center}
\end{figure}

Consider for example $\sigma$-meson exchange. The relevant diagram is illustrated in Fig.~3.
Effective interactions for $\sigma\gamma\gamma$ and $\sigma\pi\pi$ vertices can be written as
\begin{eqnarray}
\mathcal{L}_{\sigma\gamma\gamma}&=& e^2\, g_{\sigma\gamma\gamma}\, F_{\mu\nu}F^{\mu\nu}\, \sigma, \\
\mathcal{L}_{\sigma\pi\pi}&=&g_{\sigma\pi\pi}\, \vec{\pi}\cdot\vec{\pi}\, \sigma \, .
\end{eqnarray}
According to the linear sigma model~\cite{GellMann:1960np} and the quark model calculation in Ref.~\cite{Faessler:2003yf}, the coefficients $g_{\sigma\pi\pi}$ and $g_{\sigma\gamma\gamma}$ can be written as 
\begin{eqnarray}\label{ce}
g_{\sigma\gamma\gamma}\approx\frac{5}{72\, \pi^2\,F_\pi},~~~~~g_{\sigma\pi\pi}=\frac{m_\sigma^2-M_\pi^2}{2\, F_\pi}\approx\frac{m_\sigma^2}{2\, F_\pi}.
\end{eqnarray}
With a simple calculation, one obtains a magnetic polarizability contribution of
\begin{eqnarray}\label{sigma}
\beta_\pi^\sigma
&=&\frac{4\alpha}{M_\pi}\,\frac{g_{\sigma\gamma\gamma}\,g_{\sigma\pi\pi}}{m_\sigma^2} 
= \frac{5\alpha}{36\, \pi^2\, M_\pi\, F_\pi^2} = \frac{a_{-1}^\sigma}{M_\pi} \, ,
\end{eqnarray}
thus generating a leading $1/{M_\pi}$ contribution to the $\pi^0$ magnetic polarizability at
tree level with
\begin{eqnarray}\label{sigma_coeff}
a_{-1}^\sigma = 4.7 \times 10^{-4}\ \mbox{fm}^2 \, .
\end{eqnarray}
\begin{figure}[t!]
\begin{center}
\includegraphics[width=\columnwidth]{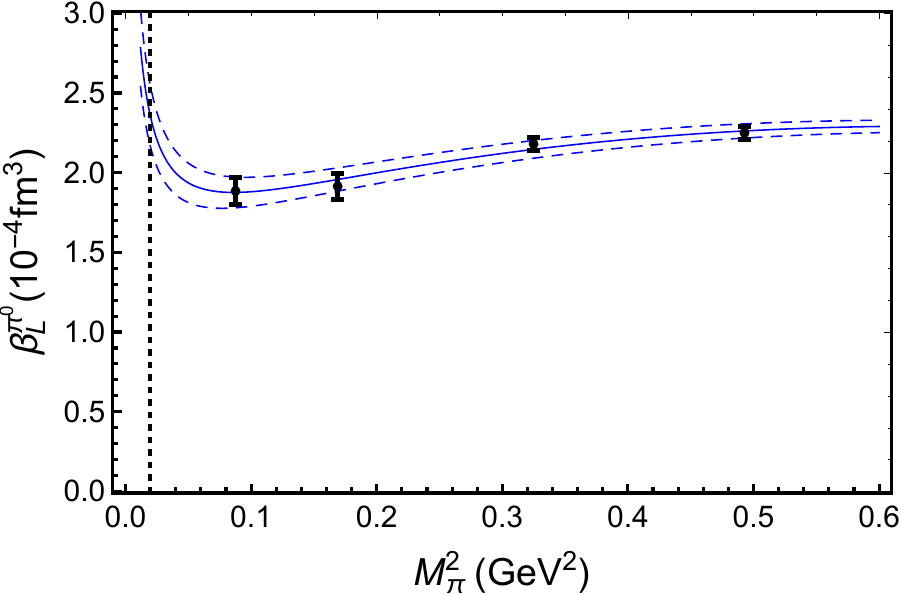}
\caption{A description of lattice QCD results \cite{Bignell:2020dze} (black points)
  for the magnetic polarizability of the neutral pion, $\beta^{\pi^0}_L$, in terms of the leading
  tree-level terms of chiral effective field theory.  The solid curve indicates the fit and the
  dot-dashed curves indicate the uncertainty associated with the statistical uncertainties of the
  lattice results.  The vertical dotted line indicates the physical point.}
\label{figN}
\end{center}
\end{figure}

While the consideration of $\sigma$-meson exchange in this manner is somewhat phenomenological, its
consideration admits a tree-level contribution proportional to $1/M_\pi$ that should be taken into
account in fitting the results from lattice QCD calculations.  
In summary, the tree-level parameterization of Eq.~(\ref{eq:tree}) is used to describe the results
of lattice QCD.  The coefficients $a_{-1}$, $a_1$ and $a_3$ are determined by fitting results from
lattice QCD \cite{Bignell:2020dze}.  
With the lattice QCD results described, one can them proceed
to include the missing contributions such as that of Eq.~(\ref{eq:leading}).

\section{Numerical Results}

A description of the lattice QCD results obtained in Ref.~\cite{Bignell:2020dze} for the magnetic
polarizability of the neutral pion, $\beta^{\pi^0}_L$, in terms of the leading tree-level terms of
Eq.~(\ref{eq:tree}) is presented in Fig.~4.  The lattice QCD results are described very
well by the tree-level contributions. The parameters obtained in the fit are
\begin{subequations}
\begin{eqnarray}
a_{-1}&=&+1.34\times10^{-4}fm^2, \\
a_1   &=&+6.85\times10^{-5}fm^4, \\
a_3   &=&-1.22\times10^{-6}fm^6 \, .
\end{eqnarray}
\label{eq:fit}
\end{subequations}
We note the leading coefficient is similar in scale to the model estimate of
Eq.~(\ref{sigma_coeff}) but suggests $\sigma$ exchange contributions are smaller
than estimated in the model.

With the fit parameters constrained, we can proceed to model the missing loop contributions
associated with diagrams (e) through (h) of Fig.~2 and thus predict the full QCD result for
$\beta_{\pi^0}$.

The correction to the leading coefficient, $a_{-1}$, is straight forward.  As explained in the
discussion of Fig.~2, the loop
contribution of Fig.~1 cannot contribute in the contemporary lattice QCD results under
consideration.  Thus in correcting for the missing contribution, we draw on Eq.~(\ref{eq:leading})
and transform
\begin{equation}
a_{-1} \to a_{-1} + \frac{\alpha}{96\, \pi^2\, F_\pi^2} \, .
\label{eq:bm1corr}
\end{equation}

To correct $a_1$ we draw on the the two-loop chiral perturbation theory calculations of
Refs.~Bellucci {\em et al.}~\cite{Bellucci:1994eb} and more recently Gasser {\it et
  al.}~\cite{Gasser:2005ud}. At this order one obtains the same leading model-independent term from
the one-loop contribution in Eq.~(\ref{eq:leading}).  The two-loop contribution introduces terms at order $\mpi$ and
non-analytic terms involving $\mpi \log \mpi$ and $\mpi \log^2 \mpi$ . The typical radius of
convergence of chiral perturbation theory is $\sim 2 \mpi$, which means that it should be
reasonable to draw from the two-loop expression of Ref.~\cite{Gasser:2005ud} to correct the
magnetic polarizability at the physical pion mass.  

Following the notation and associated values provided in Ref.~\cite{Gasser:2005ud}
\begin{equation}
\beta_{\rm 2loops} = \frac{\alpha}{32\, \pi^2\, F_\pi^2\, \mpi } 
\left [ \frac{1}{3} + 
\frac{\mpi^2 \, (d_{1+} - d_{1-})}{16\, \pi^2\, F_\pi^2} + 
{\cal O}(\mpi^4)  \right ] \, ,
\label{eq:2loops}
\end{equation}
with 
\begin{eqnarray}
d_{1+} &=& 8\,b^r  - \frac{1}{648}\,\left ( 144\,l\,(l+2\,\bar l_2) \right .
\nonumber \\
&&                  \left .+96\,l+288\,\bar l_2+ 113 + \Delta_+  \right )\,,
\nonumber \\
d_{1-} &=& a_1^r+8\,b^r +\frac{1}{648}\,\left(144\,l\,(3\,\bar l_\Delta-1) \right .
\nonumber \\
&&         \left . +36\,(8\,\bar l_1-3\,\bar l_3-12\,\bar l_4
                   +12\,\bar l_\Delta) +43 +\Delta_-\right) \,,
\nonumber \\
\Delta_+ &=& 13643-1395\,\pi^2\,,\quad\Delta_- = - 3559 + 351\,\pi^2\,.
\end{eqnarray} 
where 
$l \equiv \ln \left ( {\mpi^2}/{\mu^2} \right ),$
$\overline l_i$ are scale-independent low-energy couplings (LECs) defined in Eqs.~(3.8) and (3.9) of
Ref.~\cite{Gasser:2005ud} associated with divergences at order $p^4$ and,
$a_1^r$ and $b^r$ are low-energy couplings associated with divergences at order $p^6$, defined in
Eqs.~(3.10) and (3.11) of Ref.~\cite{Gasser:2005ud}.
The scale $\mu$ is taken to be the rho-meson mass, $\mu = M_\rho = 0.770$ GeV.  
The uncertainty in the
values of these parameters generates an uncertainty in the magnetic polarizability that will
contribute in our systematic uncertainty analysis.

The contributions from the LECs
of ${\mathcal L}_6$ contained in the coupling $a_1^r$ are associated with 
short-distance physics and therefore can have overlap with the lattice simulation results.  We proceed by replacing
the fit coefficient with the result from Eq.~(\ref{eq:2loops})
\begin{equation}
a_1 \to \frac{\alpha}{2}\, \frac{-a_1^r}{\left ( 16\, \pi^2\, F_\pi^2 \right   )^2}  \, .
\label{eq:b1corr}
\end{equation}
The remaining logarithmic terms of Eq.~(\ref{eq:2loops}) derived in the two-loop calculation are
added.  However, we also use these contributions as systematic uncertainty, both as a
measure of the possible contributions from terms of higher-order in the chiral expansion and to
account for any overlap with contributions already contained in the lattice QCD simulations.  In
summary, we model the full QCD magnetic polarizability of the neutral pion as
\begin{eqnarray}
\beta_{\pi^0}^\text{QCD}&=&
\left ( a_{-1} + \frac{\alpha}{96\, \pi^2\, F_\pi^2} \right ) \, \frac{1}{\mpi} 
\nonumber \\
&&+ \frac{\alpha}{2}\, \frac{d_{1+} - d_{1-}}{\left ( 16\, \pi^2\, F_\pi^2 \right   )^2}\, M_\pi 
+ a_3 \, M_\pi^3 \,.
\label{eq:QCD}
\end{eqnarray}

The final full-QCD prediction for the magnetic polarizability of the neutral pion is shown in
Fig.~5. There we show the original fit to the lattice QCD results and the full QCD prediction of
Eq.~(\ref{eq:QCD}) for $0 \le \mpi^2 \le 2\,\mpi^{2\,\rm Phys}$ where Eq.~(\ref{eq:QCD}) is
expected to display reasonable convergence.

Table \ref{tab:cont} provides the contributions of terms considered in Eqs.~(\ref{fit0}) and (\ref{eq:QCD}).
Here one observes the leading contribution of Eq.~(\ref{eq:QCD}) dominates the full result.
Similarly, the correction applied at order $\mpi$ in Eq.~(\ref{eq:b1corr}) is relatively small.

\begin{table*}[t] 
\caption{Contributions of terms considered in Eqs.~(\ref{fit0}) and (\ref{eq:QCD}) for the neutral
  pion magnetic polarizability in the standard units of $\times 10^{-4}$ fm$^3$.}
\label{tab:cont}
\begin{ruledtabular}
\begin{tabular}{lcc}
  Description  &Term  &Value ($\times 10^{-4}$ fm$^3$) \\
  \noalign{\smallskip}
  \hline
  \noalign{\smallskip}
  Full QCD Prediction &Eq.~(\ref{eq:QCD}) & 3.44 \\
  \noalign{\smallskip}
  \hline
  \noalign{\smallskip}
  Leading term of Eq.(\ref{eq:QCD}) &$\displaystyle \left ( a_{-1} + \frac{\alpha}{96\, \pi^2\, F_\pi^2} \right ) \, \frac{1}{\mpi}$ & 2.38\\
  \noalign{\smallskip}
  Leading one-loop contribution     &$\displaystyle \frac{\alpha}{96\, \pi^2\, F_\pi^2} \, \frac{1}{\mpi}$                           & 0.50\\
  \noalign{\smallskip}
  Leading term of Eq.~(\ref{fit0})  &$\displaystyle a_{-1} \, \frac{1}{\mpi}$                                                    &1.88 \\
  \noalign{\smallskip}
  \hline
  \noalign{\smallskip}
  Order $\mpi$ term of Eq.(\ref{eq:QCD})           &$\displaystyle \frac{\alpha}{2}\, \frac{d_{1+} - d_{1-}}{\left ( 16\, \pi^2\, F_\pi^2 \right   )^2}\, M_\pi$                         &1.06 \\
  \noalign{\smallskip}
  Order $\mpi$ correction of Eq.~(\ref{eq:b1corr}) &$\displaystyle \left ( \frac{\alpha}{2}\, \frac{d_{1+} - d_{1-}}{\left ( 16\, \pi^2\, F_\pi^2 \right   )^2} - a_1 \right ) \, M_\pi$ &0.58 \\
  \noalign{\smallskip}
  \hline
  \noalign{\smallskip}
  Order $\mpi^3$ term of Eq.(\ref{eq:QCD}) &$\displaystyle a_3 \, M_\pi^3$ & -0.004\\
\end{tabular}
\end{ruledtabular}
\end{table*}

At the physical pion mass, we find $\beta_{\pi^0} = 3.44(19)(37)
\times 10^{-4}$ fm$^3$, where the first uncertainty tems from the statistical error from fitting
the lattice QCD results and the second uncertainty is systematic as described above.
The experimental value of
$\beta_{\pi^0}^{\rm Expt} = 1.29(1.10) \times 10^{-4}$ fm$^3$ is from Ref.~\cite{Filkov:1998rwz}.  It
was determined by fitting the cross section for $\gamma\gamma\rightarrow \pi^0 \pi^0$.  Our
prediction is just above the $1 \sigma$ error bound of this experimental measurement.
\begin{figure}[t!]
\begin{center}
\includegraphics[width=\columnwidth]{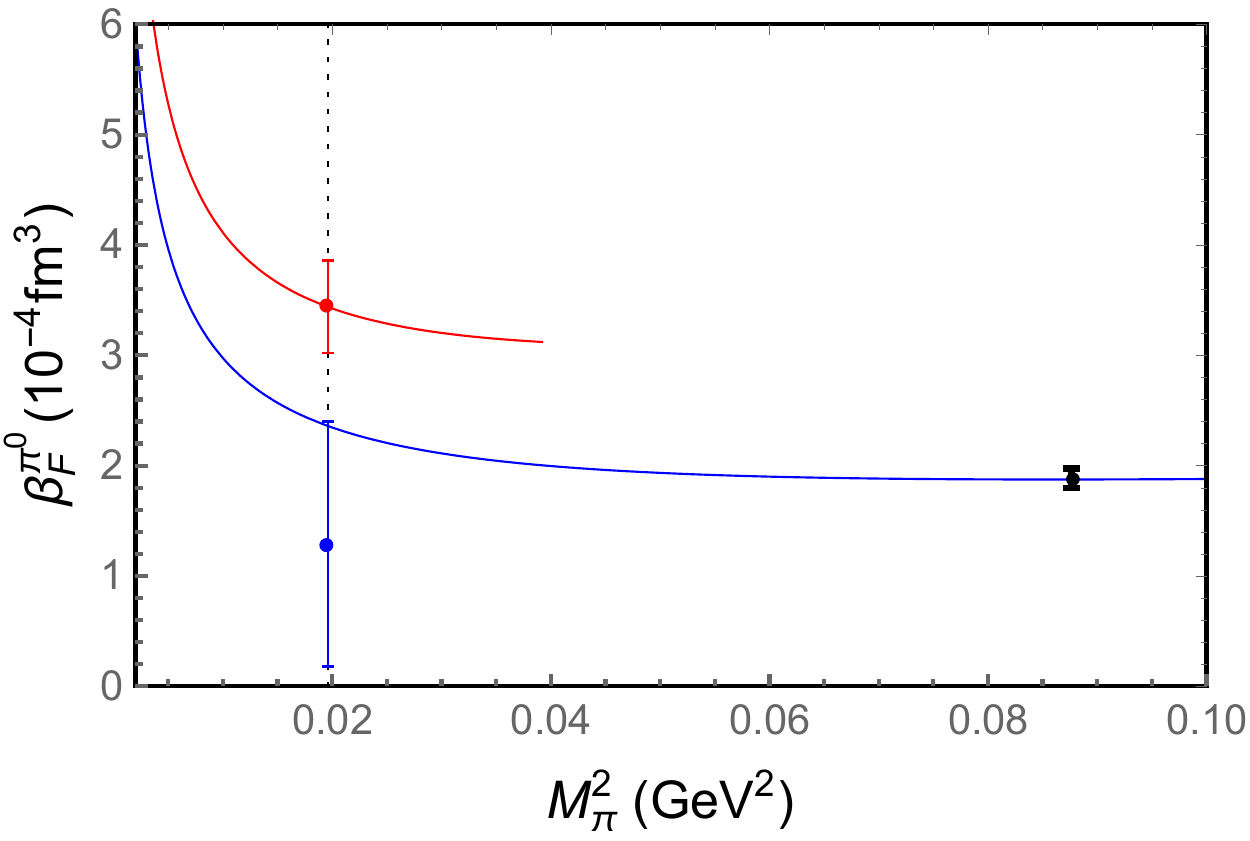}
\caption{The full QCD prediction for the magnetic polarizability of the neutral pion
  $\beta_{\pi^0}$ (red curve).  The previous fit (blue curve) of the lattice QCD simulation results
  (black points) has been corrected to incorporate pion-loop contributions absent in the current
  simulation results (red curve).  The experimental measurement (blue point) and our corresponding
  prediction (red point) are plotted at the physical pion mass.  Theoretical uncertainties include
  the statistical error from fitting the lattice QCD results and systematic uncertainties as
  described in the text.}
\label{fig:final0}
\end{center}
\end{figure}

\section{Summary}

In this paper, we have investigated the magnetic polarizability of the neutral pion based upon an
analysis of recent lattice QCD simulations at a range of quark masses.  The pion-photon scattering
amplitude is first considered at the one-loop level in partially-quenched chiral perturbation
theory.  There, the structure of the four-pion vertex causes the sea-quark loop contributions to
the magnetic polarizability at one loop to vanish. Thus the fact that the lattice simulations are
electro-quenched has no impact on the leading one-loop-contributions to the magnetic
polarizability.  

The origin of the one-loop contributions is shown to be associated with the quark-annihilation
contractions of the quark field operators of the neutral-pion interpolating fields.  As these
contributions have yet to be considered in lattice QCD, the results from contemporary calculations
are associated with tree-level terms at this order.

By considering the relationship between the Compton amplitude and the magnetic polarizability a
leading tree-level contribution proportional to $1/M_\pi$ was motivated, enabling a good
characterisation of the lattice simulation results.  The phenomenology of $\sigma$-meson exchange
provides a specific model for generating such tree-level behavior.

The full QCD result is obtained by drawing on two-loop results from chiral perturbation theory.
Because the lattice calculation does not include quark-annihilation loop contributions we are free
to add the leading contribution from the two-loop result of chiral perturbation theory with no
issue of double counting.  While LEC contributions proportional to $\mpi$ associated with
short-distance physics are used to replace the lattice fit parameter, long-distance physics
generating chiral logarithms are added to the lattice simulation results.  Our final result is
illustrated in Fig.~\ref{fig:final0}.

Our prediction for the magnetic polarizability of the neutral pion is $\beta_{\pi^0}=3.44(19)^{\rm
  stat}(37)^{\rm syst}\times 10^{-4}$ fm$^3$, just above the $1 \sigma$ error bound of the
experimental measurement.  As the experimental uncertainty is reduced in future experiments we
anticipate a significant increase in the central value.

Future research will focus on the inclusion of the quark-annihilation loop contributions in lattice
QCD. As these results become available, our fit functions will be modified to include finite-volume
effects and enable corrections to infinite volume.  In this case no modeling of the leading
contributions will be required, thus providing more robust predictions.

It will also be important to bring the techniques of partially-quenched chiral perturbation theory
to the two-loop calculation to disclose the role of sea-quark loop contributions.  While
incorporating the effects of the quark charges in lattice QED+QCD simulations is now well
established \cite{Borsanyi:2013lga,Horsley:2015eaa}, important correlations exploited in extracting
the small energy shifts relevant to polarizabilities will be lost.  This presents a formidable
challenge to calculating sea-quark-loop contributions to magnetic polarizabilities from the first
principles of QCD.

\section*{Acknowledgement}
This research was supported with supercomputing resources provided by the Phoenix HPC service at
the University of Adelaide. This research was undertaken with the assistance of resources from the
National Computational Infrastructure (NCI), provided through the National Computational Merit
Allocation Scheme, and supported by the Australian Government through Grants No.~LE190100021,
LE160100051 and the University of Adelaide Partner Share.  This research was supported by the
Australian Research Council through ARC Discovery Project Grants Nos. DP150103101 and DP180100497
(A.W.T) and DP150103164 and DP190102215 (D.B.L), and by the National Natural Sciences Foundations of China under the grant No. 11975241.

\bibliographystyle{utphys}
\bibliography{ref}

\providecommand{\href}[2]{#2}\begingroup\raggedright\begin{thebibliography}{10}

\bibitem{Aibergenov:1986gi}
T.~A. Aibergenov, P.~S. Baranov, O.~D. Beznisko, S.~N. Cherepniya, L.~V.
  Filkov, A.~A. Nafikov, A.~I. Osadchii, V.~G. Raevsky, L.~N. Shtarkov, and
  E.~I. Tamm, ``{Radiative Photoproduction of Pions and Pion Compton
  Scattering},''
\href{http://dx.doi.org/10.1007/BF01797507}{{\em Czech. J. Phys.} {\bfseries
  B36} (1986) 948--951}.

\bibitem{Antipov:1982kz}
{\relax Yu}.~M. Antipov {\em et~al.}, ``{Measurement of pi- Meson
  Polarizability in Pion Compton Effect},''
\href{http://dx.doi.org/10.1016/0370-2693(83)91195-4}{{\em Phys. Lett.}
  {\bfseries 121B} (1983) 445--448}.

\bibitem{Antipov:1984ez}
{\relax Yu}.~M. Antipov {\em et~al.}, ``{Experimental Evaluation of the Sum of
  the Electric and Magnetic Polarizabilities of Pions},''
\href{http://dx.doi.org/10.1007/BF01551790}{{\em Z. Phys.} {\bfseries C26}
  (1985) 495}.

\bibitem{Boyer:1990vu}
J.~Boyer {\em et~al.}, ``{Two photon production of pion pairs},''
\href{http://dx.doi.org/10.1103/PhysRevD.42.1350}{{\em Phys. Rev.} {\bfseries
  D42} (1990) 1350--1367}.

\bibitem{Marsiske:1990hx}
{\bfseries Crystal Ball} Collaboration, H.~Marsiske {\em et~al.}, ``{A
  Measurement of $\pi^0 \pi^0$ Production in Two Photon Collisions},''
\href{http://dx.doi.org/10.1103/PhysRevD.41.3324}{{\em Phys. Rev.} {\bfseries
  D41} (1990) 3324}.

\bibitem{Filkov:2005ccw}
L.~V. Fil'kov and V.~L. Kashevarov, ``{Determination of pi0 meson quadrupole
  polarizabilities from the process gamma gamma ---> pi0 pi0},''
  \href{http://dx.doi.org/10.1103/PhysRevC.72.035211}{{\em Phys. Rev.}
  {\bfseries C72} (2005) 035211},
\href{http://arxiv.org/abs/nucl-th/0505058}{{\ttfamily arXiv:nucl-th/0505058
  [nucl-th]}}.

\bibitem{Bernard:1988wi}
V.~Bernard and D.~Vautherin, ``{Electromagnetic Polarizabilities of
  Pseudoscalar Goldstone Bosons},''
\href{http://dx.doi.org/10.1103/PhysRevD.40.1615}{{\em Phys. Rev.} {\bfseries
  D40} (1989) 1615}.

\bibitem{Ivanov:1991kw}
M.~A. Ivanov and T.~Mizutani, ``{Pion and kaon polarizabilities in the quark
  confinement model},''
\href{http://dx.doi.org/10.1103/PhysRevD.45.1580}{{\em Phys. Rev.} {\bfseries
  D45} (1992) 1580--1601}.

\bibitem{Bernard:1992mp}
V.~Bernard, A.~A. Osipov, and U.~G. Meissner, ``{Consistent treatment of the
  bosonized Nambu-Jona-Lasinio model},''
\href{http://dx.doi.org/10.1016/0370-2693(92)91309-W}{{\em Phys. Lett.}
  {\bfseries B285} (1992) 119--125}.

\bibitem{Wilmot:2002gg}
C.~A. Wilmot and R.~H. Lemmer, ``{Electric and magnetic polarizability of
  Goldstone pions to subleading O(Nc-1) in the bosonized Nambu-Jona-Lasinio
  model},''
\href{http://dx.doi.org/10.1103/PhysRevC.65.035206}{{\em Phys. Rev.} {\bfseries
  C65} (2002) 035206}.

\bibitem{Bijnens:1987dc}
J.~Bijnens and F.~Cornet, ``{Two Pion Production in Photon-Photon
  Collisions},''
\href{http://dx.doi.org/10.1016/0550-3213(88)90032-6}{{\em Nucl. Phys.}
  {\bfseries B296} (1988) 557--568}.

\bibitem{Bellucci:1994eb}
S.~Bellucci, J.~Gasser, and M.~Sainio, ``{Low-energy photon-photon collisions
  to two loop order},''
  \href{http://dx.doi.org/10.1016/0550-3213(94)90566-5}{{\em Nucl. Phys. B}
  {\bfseries 423} (1994) 80--122},
  \href{http://arxiv.org/abs/hep-ph/9401206}{{\ttfamily arXiv:hep-ph/9401206}}.
  [Erratum: Nucl.Phys.B 431, 413--414 (1994)].

\bibitem{Burgi:1996qi}
U.~Burgi, ``{Charged pion pair production and pion polarizabilities to two
  loops},'' \href{http://dx.doi.org/10.1016/0550-3213(96)00454-3}{{\em Nucl.
  Phys.} {\bfseries B479} (1996) 392--426},
\href{http://arxiv.org/abs/hep-ph/9602429}{{\ttfamily arXiv:hep-ph/9602429
  [hep-ph]}}.

\bibitem{Gasser:2005ud}
J.~Gasser, M.~A. Ivanov, and M.~E. Sainio, ``{Low-energy photon-photon
  collisions to two loops revisited},''
  \href{http://dx.doi.org/10.1016/j.nuclphysb.2005.09.010}{{\em Nucl. Phys.}
  {\bfseries B728} (2005) 31--54},
\href{http://arxiv.org/abs/hep-ph/0506265}{{\ttfamily arXiv:hep-ph/0506265
  [hep-ph]}}.

\bibitem{Moinester:2019sew}
M.~Moinester and S.~Scherer, ``{Compton Scattering off Pions and
  Electromagnetic Polarizabilities},''
  \href{http://dx.doi.org/10.1142/S0217751X19300084}{{\em Int. J. Mod. Phys.}
  {\bfseries A34} no.~16, (2019) 1930008},
\href{http://arxiv.org/abs/1905.05640}{{\ttfamily arXiv:1905.05640 [hep-ph]}}.

\bibitem{Filkov:1982cx}
L.~V. Filkov, I.~Guiasu, and E.~E. Radescu, ``{Pion Polarizabilities From
  Backward and Fixed $u$ Sum Rules},''
\href{http://dx.doi.org/10.1103/PhysRevD.26.3146}{{\em Phys. Rev.} {\bfseries
  D26} (1982) 3146}.

\bibitem{Donoghue:1993kw}
J.~F. Donoghue and B.~R. Holstein, ``{Photon-photon scattering, pion
  polarizability and chiral symmetry},''
  \href{http://dx.doi.org/10.1103/PhysRevD.48.137}{{\em Phys. Rev.} {\bfseries
  D48} (1993) 137--146},
\href{http://arxiv.org/abs/hep-ph/9302203}{{\ttfamily arXiv:hep-ph/9302203
  [hep-ph]}}.

\bibitem{Bernard:1988gp}
V.~Bernard, B.~Hiller, and W.~Weise, ``{Pion Electromagnetic Polarizability and
  Chiral Models},''
\href{http://dx.doi.org/10.1016/0370-2693(88)90391-7}{{\em Phys. Lett.}
  {\bfseries B205} (1988) 16--21}.

\bibitem{Burkardt:1996vb}
M.~Burkardt, D.~B. Leinweber, and X.-m. Jin, ``{Background field formalism in
  quantum systems},''
  \href{http://dx.doi.org/10.1016/0370-2693(96)00881-7}{{\em Phys. Lett. B}
  {\bfseries 385} (1996) 52--56},
  \href{http://arxiv.org/abs/hep-ph/9604450}{{\ttfamily arXiv:hep-ph/9604450}}.

\bibitem{Primer:2013pva}
T.~Primer, W.~Kamleh, D.~Leinweber, and M.~Burkardt, ``{Magnetic properties of
  the nucleon in a uniform background field},''
  \href{http://dx.doi.org/10.1103/PhysRevD.89.034508}{{\em Phys. Rev. D}
  {\bfseries 89} no.~3, (2014) 034508},
  \href{http://arxiv.org/abs/1307.1509}{{\ttfamily arXiv:1307.1509 [hep-lat]}}.

\bibitem{Bignell:2018acn}
R.~Bignell, J.~Hall, W.~Kamleh, D.~Leinweber, and M.~Burkardt, ``{Neutron
  magnetic polarizability with Landau mode operators},''
  \href{http://dx.doi.org/10.1103/PhysRevD.98.034504}{{\em Phys. Rev.}
  {\bfseries D98} no.~3, (2018) 034504},
\href{http://arxiv.org/abs/1804.06574}{{\ttfamily arXiv:1804.06574 [hep-lat]}}.

\bibitem{Bignell:2020xkf}
R.~Bignell, W.~Kamleh, and D.~Leinweber, ``{Magnetic polarizability of the
  nucleon using a Laplacian mode projection},''
  \href{http://dx.doi.org/10.1103/PhysRevD.101.094502}{{\em Phys. Rev. D}
  {\bfseries 101} no.~9, (2020) 094502},
  \href{http://arxiv.org/abs/2002.07915}{{\ttfamily arXiv:2002.07915
  [hep-lat]}}.

\bibitem{Luschevskaya:2014lga}
E.~Luschevskaya, O.~Solovjeva, O.~Kochetkov, and O.~Teryaev, ``{Magnetic
  polarizabilities of light mesons in $SU(3)$ lattice gauge theory},''
  \href{http://dx.doi.org/10.1016/j.nuclphysb.2015.07.023}{{\em Nucl. Phys. B}
  {\bfseries 898} (2015) 627--643},
  \href{http://arxiv.org/abs/1411.4284}{{\ttfamily arXiv:1411.4284 [hep-lat]}}.

\bibitem{Luschevskaya:2015cko}
E.~Luschevskaya, O.~Solovjeva, and O.~Teryaev, ``{Magnetic polarizability of
  pion},'' \href{http://dx.doi.org/10.1016/j.physletb.2016.08.054}{{\em Phys.
  Lett. B} {\bfseries 761} (2016) 393--398},
  \href{http://arxiv.org/abs/1511.09316}{{\ttfamily arXiv:1511.09316
  [hep-lat]}}.

\bibitem{Bignell:2019vpy}
R.~Bignell, W.~Kamleh, and D.~Leinweber, ``{Pion in a uniform background
  magnetic field with clover fermions},''
  \href{http://dx.doi.org/10.1103/PhysRevD.100.114518}{{\em Phys. Rev.}
  {\bfseries D100} no.~11, (2020) 114518},
  \href{http://arxiv.org/abs/1910.14244}{{\ttfamily arXiv:1910.14244
  [hep-lat]}}.
[Phys. Rev.D100,114518(2019)].

\bibitem{Ding:2020hxw}
H.-T. Ding, S.-T. Li, A.~Tomiya, X.-D. Wang, and Y.~Zhang, ``{Chiral properties
  of (2+1)-flavor QCD in strong magnetic fields at zero temperature},''
  \href{http://arxiv.org/abs/2008.00493}{{\ttfamily arXiv:2008.00493
  [hep-lat]}}.

\bibitem{Bignell:2020dze}
R.~Bignell, W.~Kamleh, and D.~Leinweber, ``{Pion magnetic polarisability using
  the background field method},''
\href{http://arxiv.org/abs/2005.10453}{{\ttfamily arXiv:2005.10453 [hep-lat]}}.

\bibitem{Hall:2013dva}
J.~M.~M. Hall, D.~B. Leinweber, and R.~D. Young, ``{Finite-volume and partial
  quenching effects in the magnetic polarizability of the neutron},''
  \href{http://dx.doi.org/10.1103/PhysRevD.89.054511}{{\em Phys. Rev.}
  {\bfseries D89} no.~5, (2014) 054511},
\href{http://arxiv.org/abs/1312.5781}{{\ttfamily arXiv:1312.5781 [hep-lat]}}.

\bibitem{Hu:2007ts}
J.~Hu, F.-J. Jiang, and B.~C. Tiburzi, ``{Pion Polarizabilities and Volume
  Effects in Lattice QCD},''
  \href{http://dx.doi.org/10.1103/PhysRevD.77.014502}{{\em Phys. Rev. D}
  {\bfseries 77} (2008) 014502},
  \href{http://arxiv.org/abs/0709.1955}{{\ttfamily arXiv:0709.1955 [hep-lat]}}.

\bibitem{Bernard:1992mk}
C.~W. Bernard and M.~F.~L. Golterman, ``{Chiral perturbation theory for the
  quenched approximation of QCD},''
  \href{http://dx.doi.org/10.1103/PhysRevD.46.853}{{\em Phys. Rev.} {\bfseries
  D46} (1992) 853--857},
\href{http://arxiv.org/abs/hep-lat/9204007}{{\ttfamily arXiv:hep-lat/9204007
  [hep-lat]}}.

\bibitem{Leinweber:2002qb}
D.~B. Leinweber, ``{Quark contributions to baryon magnetic moments in full,
  quenched and partially quenched QCD},''
  \href{http://dx.doi.org/10.1103/PhysRevD.69.014005}{{\em Phys. Rev.}
  {\bfseries D69} (2004) 014005},
\href{http://arxiv.org/abs/hep-lat/0211017}{{\ttfamily arXiv:hep-lat/0211017
  [hep-lat]}}.

\bibitem{GellMann:1960np}
M.~Gell-Mann and M.~Levy, ``{The axial vector current in beta decay},''
\href{http://dx.doi.org/10.1007/BF02859738}{{\em Nuovo Cim.} {\bfseries 16}
  (1960) 705}.

\bibitem{Faessler:2003yf}
A.~Faessler, T.~Gutsche, M.~A. Ivanov, V.~E. Lyubovitskij, and P.~Wang, ``{Pion
  and sigma meson properties in a relativistic quark model},''
  \href{http://dx.doi.org/10.1103/PhysRevD.68.014011}{{\em Phys. Rev.}
  {\bfseries D68} (2003) 014011},
\href{http://arxiv.org/abs/hep-ph/0304031}{{\ttfamily arXiv:hep-ph/0304031
  [hep-ph]}}.

\bibitem{Filkov:1998rwz}
L.~V. Fil'kov and V.~L. Kashevarov, ``{Compton scattering on the charged pion
  and the process gamma gamma ---> pi0 pi0},''
  \href{http://dx.doi.org/10.1007/s100500050287}{{\em Eur. Phys. J.} {\bfseries
  A5} (1999) 285--292},
\href{http://arxiv.org/abs/nucl-th/9810074}{{\ttfamily arXiv:nucl-th/9810074
  [nucl-th]}}.

\bibitem{Borsanyi:2013lga}
{\bfseries Budapest-Marseille-Wuppertal} Collaboration, S.~Borsanyi {\em
  et~al.}, ``{Isospin splittings in the light baryon octet from lattice QCD and
  QED},'' \href{http://dx.doi.org/10.1103/PhysRevLett.111.252001}{{\em Phys.
  Rev. Lett.} {\bfseries 111} no.~25, (2013) 252001},
  \href{http://arxiv.org/abs/1306.2287}{{\ttfamily arXiv:1306.2287 [hep-lat]}}.

\bibitem{Horsley:2015eaa}
R.~Horsley {\em et~al.}, ``{Isospin splittings of meson and baryon masses from
  three-flavor lattice QCD + QED},''
  \href{http://dx.doi.org/10.1088/0954-3899/43/10/10LT02}{{\em J. Phys. G}
  {\bfseries 43} no.~10, (2016) 10LT02},
  \href{http://arxiv.org/abs/1508.06401}{{\ttfamily arXiv:1508.06401
  [hep-lat]}}.

\end{thebibliography}\endgroup

\end{document}